\documentclass{aa}  

\usepackage{graphicx}
\graphicspath{{figures/}}

\newcommand{\omc}{${\rm O}-{\rm C}$}
\newcommand{\omcl}{{\rm O}-{\rm C}}
\newcommand{\msol}{~\ensuremath{\mathrm{M}_\odot}}
\newcommand{\rsol}{~\ensuremath{\mathrm{R}_\odot}}

\usepackage{txfonts}
\usepackage{bm}
\usepackage[colorlinks,allcolors=blue]{hyperref}
\begin{document}

\title{Eclipsing time variations in close binaries\\ produced by azimuthal dynamo waves}
\titlerunning{ETVs from ADWs}

\author{Felipe H. Navarrete\inst{1,2}
          \and
          Dominik R.G. Schleicher\inst{3}
          \and
          Petri J. K\"apyl\"a\inst{4}
          \and
          Marcel V\"olschow\inst{5,6}
          }
   \institute{
            Institute of Space Sciences, (ICE-CSIC), Campus UAB, Carrer de
            Can Magrans s/n, 08193 Barcelona, Spain\\
            \email{navarrete@ice.csic.es}
         \and
             GERICS, Helmholtz-Zentrum Hereon, Fischertwiete 1, 20095 Hamburg, Germany  
         \and        Dipartimento di Fisica, Sapienza Università di Roma, Piazzale Aldo Moro 5, 00185 Rome, Italy
         \and
            Institute for Solar Physics (KIS), Georges-K\"ohler-Allee 401a,
            79110 Freiburg, Germany
         \and
        Hamburg University of Applied Sciences, 
        Berliner Tor 7, 20099 Hamburg, Germany
        \and
        Centre for X-ray and Nano Science, Deutsches Elektronen-Synchrotron DESY, 
        Notkestraße 85, 22607 Hamburg, Germany
             }

   \date{Received --; accepted --}

 
  \abstract
   {The nature of eclipsing time variations (ETVs) in post-common-envelope
     binaries (PCEBs) is still unknown. Circumbinary planets routinely fail
     the test of time and the Applegate mechanism has energetic constraints
     and problems in reproducing observations.
   }
   {Based on recent analytic models of magnetically-induced ETVs and stellar dynamo
   simulations, we aim at explaining ETVs via non-axisymmetric magnetic fields that
   drift in the azimuthal direction of the star, know as azimuthal dynamo waves (ADWs).
   }
   {We implement a time-varying non-axisymmetric quadrupole moment in a binary
    system. We solve for the dynamics of the system,
    compute the resulting eclipsing times, and construct \omc\, diagrams.
    We perform several simulations with different quadrupole moment amplitudes, periods,
    stellar masses and binary separations.
   }
   {ADWs naturally give rise to characteristic shapes in the \omc\,
    diagram that resemble observations. Depending on how fast the quadrupole
    moment changes, the solutions can have a sharp decrease in \omc\, producing
    amplitudes such as the one obtained in QS~Vir, or sinusoidal-like shapes
    such as in V471~Tau or NN~Ser. We also find that the amplitude of the
    eclipsing times varies from tens to hundreds of seconds.
   }
   {ADWs offer a self-consistent explanation for ETVs as their presence is
    expected in dynamo theory. They can explain a
    variety of features in the observed O-C diagrams. As suggested by dynamo simulations,
    ADWs are easily excited in rapidly rotating stars, strongly alleviating
    energetic constrains required in the context of, for example, the Applegate
    mechanism. They produce non-axisymmetric quadrupole
    moments that in turn produce ETVs that can account for the long-term
    variation of the \omc\, diagrams. We expect in this
    case that the resulting \omc\, diagrams are not strictly periodic, unlike
    explanations based on a third body that would imply a strict periodicity
    unless additional mechanisms are being invoked.}


   \maketitle

\section{Introduction}

Post-common-envelope binaries (PCEBs) are a class of binaries which are
characterized by small binary separations, a magnetically active star,
and a primary star which is usually a white dwarf. As their name suggests, they are believed to have formed through
a common envelope (CE) stage \citep{paczynski76}. A peculiar characteristic of PCEBs
is that they show eclipsing time variations (ETVs) with quasi-periodic variations
of the order of tens to several hundred seconds
\citep{zorotovic13}. These ETVs are usually studied through the
observed-minus-calculated (\omc) diagram of the eclipsing times, which is
constructed by subtracting the expected eclipsing times from the actual
observed eclipsing time \citep[e.g.][]{qian09,beuermann10,kundra22,pulley22}.

The periodic nature of the \omc\, diagram has eluded an explanation despite
having data that span over decades for many PCEBs. Two competing scenarios,
albeit not mutually exclusive, have been proposed to explain ETVs in PCEBs. The \omc\, variations can be
produced by the light-travel-time effect (LTTE) by planets that either
formed together with the binary and have survived the common envelope phase \citep[known as first generation formation scenario: see, e.g.,][]
{voelschow16}, or formed from the ejected material of the CE
\citep[the so-called second generation formation scenario; see, e.g.,][]{schleicher14, schleicher15}.
Another explanation links the eclipsing times to the gravitational quadrupole
moment of the magnetically active star through magnetic activity
\citep{Applegate92,Lanza20}. Both scenarios, namely planets and magnetic
activity, have their own advantages and pitfalls. For example, planets can be formed in
binary stars under several different scenarios \citep{tutukov12}, but at least
first generation planets are more difficult to reconcile with observational data \citep{voelschow14,schleicher15}.
However, the planetary hypothesis has been called into question as its predictions for new eclipsing times were not consistent with the subsequent observations. In particular, \citet{pulley22} studied
seven PCEBs with ``confirmed'' circumbinary planets providing new eclipsing times
for all of them. For six out of the seven systems, previously published planetary
models failed to predict the new data points. On the other hand, the standard
mechanism that explains ETVs through magnetic activity (known as Applegate
mechanism; see Sect.~\ref{sec:review}) routinely requires energies larger than
what is provided by the active component \citep{brinkworth06,voelschow16}. Recently,
direct numerical simulations of stellar dynamos  had difficulties to reproduce
quadrupole moment variations large enough to explain ETVs with
the Applegate mechanism \citep{Navarrete20,Navarrete22,Navarrete22b}.

\citet{Lanza20}, based on the study by \citet{Applegate89}, presented a model
in which the energy constraints of the Applegate mechanism are lifted 
(see Sect.~\ref{sec:review} for a description of the mechanism).
However, this mechanism relies on two assumptions that are difficult to
reconcile. The first one is that the very compact binary system is not tidally locked; the second is that
the magnetically active star has a permanent non-axisymmetric magnetic field,
modeled as a single flux tube \citep{Lanza20}, which is frozen into the
convective zone (CZ) of the star and thus has no dynamical evolution. The tidal
synchronization
timescale can be computed as $\tau{_{\rm sync}} = f{_{\rm turb}}q^{-2}(a/R)^6$
yrs, where $f{_{\rm turb}}\sim1$, $q = m_{0}/m_{1}$ is the ratio between the
companion (white dwarf) and magnetically active star (main-sequence star),
$R$ is the radius of the main-sequence star,
and $a$ is the binary separation
\citep{Zahn77,Toledo07,Song13}. Using $q=0.5$, $a=1~\rsol$,
and $R=0.2\rsol$, which are typical parameters for PCEBs, yields
$\tau_{_{\rm sync}} \sim 62{,}000~\mathrm{yrs}$. Given this very short synchronization
timescale, it is expected that most PCEBs, if not all, have already reached tidal
synchronization. Therefore the expected degree of
asynchronism should be very small \citep{Lanza20}. Nonetheless, \citet{Lurie17} found
asynchronous close binaries with short periods that have not synchronized
their orbital and rotational periods. This could be attributed to either
young systems or complex dynamical histories like the one found in CE events.

{ More recent studies have incorporated new eclipsing times and further studied
whether magnetic activity and circumbinary planets can reproduce the observations.
\cite{Pulley2025} report 355 eclipsing times for 7 PCEBs. The authors find that
in neither of them the new observations fit the proposed planetary models,
requiring once again to change the proposed circumbinary planet parameters.
At the same time, the authors find that the energy available in the magnetically
active star to drive either the Applegate or the Lanza mechanisms is insufficient
to be the responsible of the full ETVs. A similar conclusion has been reached by \cite{Yates2026}, with the difference that the latter do find a set of systems that
fulfill the energetic requirements to reproduce the ETVs with magnetic fields.}

On the other hand, strong non-axisymmetric magnetic field
modes are commonly obtained in stellar dynamo simulations \citep{Cole14,
Viviani18,Kapyla20a,Navarrete22}, which have a time dependence
and which are often stronger than axisymmetric modes. Furthermore, they often
drift in the azimuthal direction independent of the flow. It is important to distinguish between the planetary hypothesis and
magnetically-driven quadrupole moment variations for the interpretation of \omc\,
diagrams. They might either serve as a way to study planetary formation (either
first or second generation planets), and their orbital stability in the case of
an extreme event such as the CE phase. Or if the origin is due to magnetism, such diagrams could be
used to study stellar dynamos because the shape of the \omc\, could be directly
linked to the strength, topology, and time evolution of the magnetic field in
the CZ of the magnetically active star. The correct interpretation is therefore highly relevant, and in case of specific predictions of planets, direct detection experiments should be encouraged, to either confirm or falsify the planetary hypothesis.

In this paper, we study the hypothesis of ETVs being induced by magnetic
activity in the active component of close binaries.
We use N-Body simulations of binaries, modeled as point particles, with the
main-sequence component having a quadrupole moment that rotates due to a migrating azimuthal dynamo wave (ADW; see Sect.~\ref{sec:ADWs}).

In Sect.~\ref{sec:review} we review and contrast the Applegate mechanism and the  \citet{Lanza20} mechanism, as well as ADWs as an alternative mechanism to produce ETVs. To explore the \omc\, diagrams produced by magnetically-driven quadrupole moments, we employ numerical models of binary systems described in Sect.~\ref{sec:model}. Section~\ref{sec:results} follows with the
results of our simulations.
We finally discuss our results and present our conclusions in Sect.~\ref{sec:discuss}.


\section{Theory}\label{sec:review}

Two different mechanisms to explain ETVs with stellar magnetic fields have been
identified in the literature so far. The first is the Applegate mechanism \citep{Applegate92} and the
second was recently presented by \cite{Lanza20}, based on the
earlier work of \cite{Applegate89}. We refer to the latter henceforth as 
the Applegate-Lanza mechanism. In this section we briefly review them as well
as a new scenario based on ADWs.

Generally speaking, the connection between ETVs and magnetic activity can be
seen by writing the gravitational potential of the magnetically active star
as
\begin{equation}\label{eq:gravpot}
    \phi(r) = -\frac{GM}{r} - \frac{3G}{2r^3}\sum_{ij}\frac{Q_{ij}x_{i}x_{j}}{r^2},
\end{equation}
where $G$ is the gravitational constant, $M$ is the mass of the companion star,
$r$ is the distance between the centers of the stars, $Q_{ij}$ are the
components of the quadrupole moment, and $x_{i,j}$ are Cartesian coordinates
measured from the center of the magnetically active star.
Explicitly, the components of the quadrupole moment are computed as
\begin{equation}\label{eq:quadrupole}
    Q_{ij} = I_{ij} - \frac{1}{3}\delta_{ij}{\rm Tr}\,I,
\end{equation}
where
\begin{equation}\label{eq:inertia}
    I_{ij} = \int \rho({\bf x})x_{i}x_{j}{\rm d}V,
\end{equation}
where $\delta_{ij}$ is the Kronecker delta, $I_{ij}$ are the components of the
moment of inertia, ${\rm Tr}\,I$ is its trace, and ${\bf x}=(x_{i},x_{j},x_{k})$
are Cartesian coordinates.

By changing the components of the moment of inertia (Eq.~\ref{eq:inertia}),
the quadrupole moment (Eq.~\ref{eq:quadrupole}) and the gravitational potential
(Eq.~\ref{eq:gravpot}) also change. Magnetic activity may change the flow patterns inside the star through various mechanisms, including changes in global circulation, angular momentum transport or the propagation of dynamo waves inside the star, thereby affecting the quadrupole moment and the \omc\, diagrams.

\subsection{Applegate mechanism}\label{sec:applegate}

The Applegate mechanism works with a time-dependent axisymmetric quadrupole
moment. The idea here is that during a magnetic cycle, angular momentum is
transported within the star from the deep CZ to the surface with the sub-surface
magnetic field providing the necessary torque. An increased
angular momentum in the surface layers would then make the star increase its
radius due to an increase of the centrifugal force. In this context, the
radius of the star is directly related to the eclipsing time.

By placing the origin of the Cartesian coordinate system in the center of the
magnetically active star, where the $x$-axis points towards the companion and
the $z$-axis is parallel to the stellar rotation axis, \citet{Applegate92} further
assumed that the secondary star is tidally locked. This implies that only the
$xx$-component of the quadrupole moment contributes to the gravitational
potential. Namely,
\begin{equation}
    \phi(r) = -\frac{GM}{r} - \frac{3G}{2r^3}Q_{xx}.
\end{equation}
Variations of the quadrupole moment $\Delta Q_{xx}$ induce binary period
variations \citep{Applegate92}
\begin{equation}\label{eq:Delp_Qxx}
    \frac{\Delta P}{P} = -9\left(\frac{R}{a}\right)^2\frac{\Delta Q_{xx}}{MR^2},
\end{equation}
where $\Delta P$ is the period variation between two orbital periods, $P$ is the
orbital period, $R$ is the radius of the magnetically active star, $M$ is its mass, and $a$ is the
binary separation. For a system such as NN~Ser, the required quadrupole moment variation to produce the observed relative period variations is
$7.38\times10^{39}$~kg~m$^2$. For V471~Tau, the required quadrupole moment
variation is $8.19\times10^{41}$~kg~m$^2$

The mechanism which produces $\Delta Q_{xx}$ relies on the strength of the
magnetic field. The necessary magnetic field strength to produce a given
$\Delta P$ and $P_{\rm mod}$ is given by
\begin{equation}
    B \sim \sqrt{10 \frac{GM^2}{R^4} \frac{\Delta P}{P_{\rm mod}}}\left(\frac{a}{R}\right),
\end{equation}
which for the parameters of NN~Ser gives a field strength of about 42~kG. For V471~Tau it results in 10~kG. These magnetic fields are assumed to apply a torque acting at the subsurface level, redistributing the angular momentum
between the interior part and the surface layers \citep{Applegate92}. While the
magnetic field for V471~Tau is achieved in stellar dynamo simulations, it is
not the case for NN~Ser. However, the central criticism of the Applegate
mechanism is that the necessary energy to drive it is larger than what is
available from stellar luminosity in several cases. \citet{brinkworth06} extended the Applegate
mechanism to a more general case in which the shell that surrounds the internal
(central) mass of the star is finite, and also the quadrupole moment
of this core is considered. In such a case, the necessary energy to drive
the Applegate mechanism is given by
\begin{equation}
    \Delta E = \Omega_{\rm dr}\Delta J + \frac{1}{2}\left(\frac{1}{I_{1}} + \frac{1}{I_{2}}\right)(\Delta J)^2,
\end{equation}
where $\Delta E$ is the energy change, $\Omega_{\rm dr}$ is the differential
rotation, defined as the difference in angular frequencies between the stellar
outer shell and the inner core, $\Delta J$ is the angular momentum change, and
$I_{1}$ and $I_{2}$ are
the moments of inertia of the inner core and outer shell, respectively.
For NN Ser, \citet{brinkworth06} computed that
\begin{equation}
4\times10^{40} \lesssim \Delta E \lesssim 1\times10^{42}\,{\rm erg},
\end{equation}
whereas the available energy in the secondary of NN~Ser, computed from the effective temperature, falls in the range of
\begin{equation}
    2.5\times10^{39} \leq E \leq 4.5\times10^{39}\,{\rm erg}.
\end{equation}
The required energy is more than ten times larger than what is available
from the stellar luminosity, and thus the Applegate mechanism is unable to explain the observed ETVs in NN~Ser.

A subsequent improvement to the model was put forward by \citet{Voelschow18},
who considered period modulations arising from kinetic and magnetic field
fluctuations. Only a few selected systems were able to reproduce the
expected period variation. The binaries that could host ETVs driven
by the Applegate mechanism were those with a tight binary separation of less than
$1~\rsol$ and secondary masses of $\simeq 0.35~\msol$.

\subsection{Applegate-Lanza mechanism}\label{sec:applegate-lanza}

The Applegate-Lanza mechanism presented in \citet{Lanza20} is based on two
key assumptions: (i) there is
a permanent magnetic flux tube in the convective zone of the magnetically
active star of the PCEB, and (ii) the system has not yet reached tidal
synchronization. The magnetic flux tube perturbs the density field
because in its interior the density must decrease in order to keep magnetic and
fluid pressure in balance. Because of this, a permanent non-axisymmetric
quadrupole moment contributes to the gravitational potential of the star.
Furthermore, relaxing the condition of tidal locking makes the
non-axisymmetric structure rotate in the frame of reference of the
binary. Due to the asymmetry of the density field and its rotation,
the companion star experiences a time-dependent gravitational potential.

\begin{figure}
\centering
\includegraphics[width=\columnwidth]{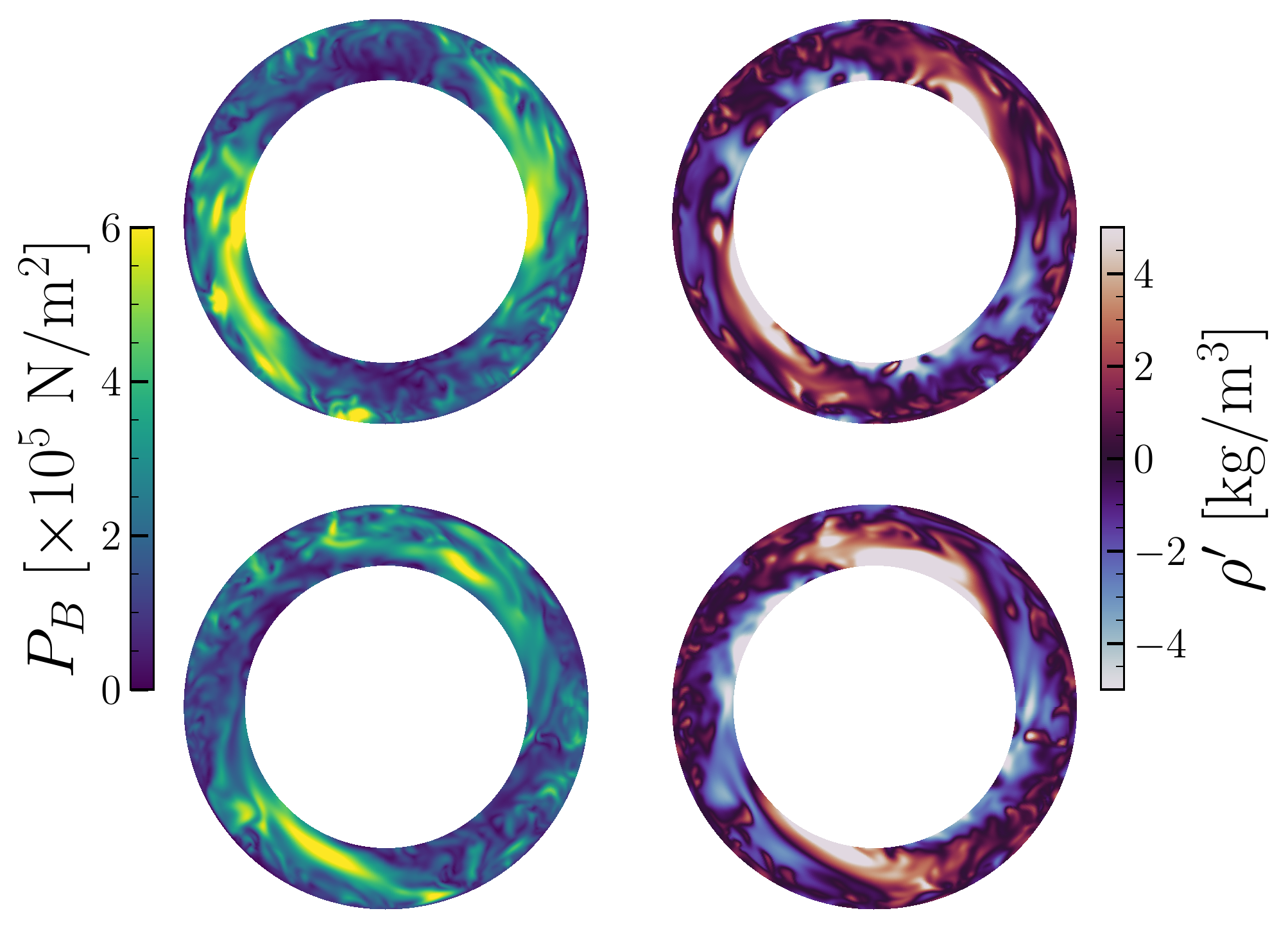}
\caption{Magnetic pressure (left panels) and non-axisymmetric density
field (right panels) from run B of \citet{Navarrete22}
at latitudes $+60^\circ$
(top panels) and $-60^\circ$ (bottom panels) from a dynamo simulation of a solar-mass star with a rotation period of $1.2$~days.
}
\label{fig:PB_rho}%
\end{figure}

\begin{figure}
\centering
\includegraphics[width=\columnwidth]{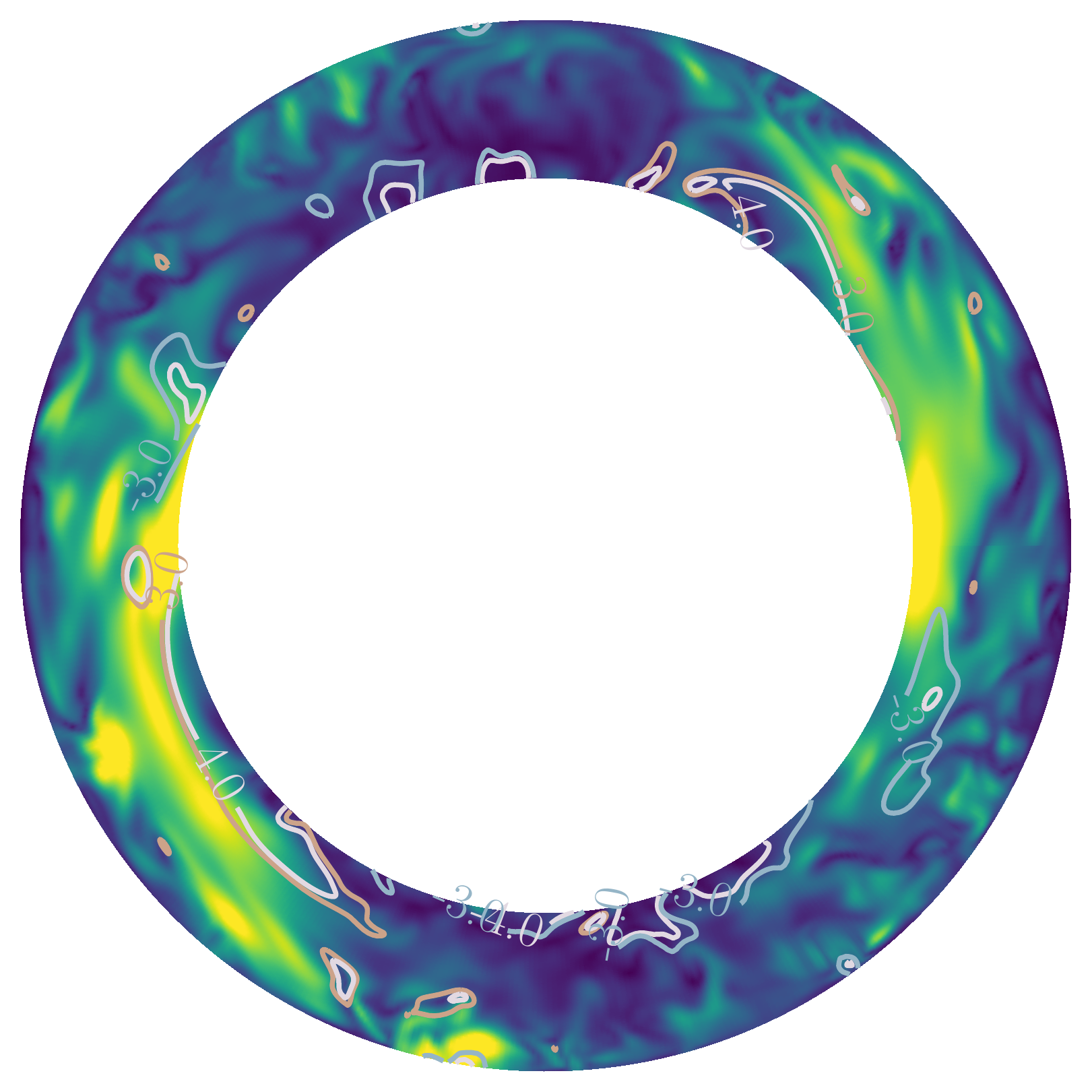}
\caption{Same as the top panels of Fig.~\ref{fig:PB_rho} but the density
field is show as isocontours at the $\pm3$ and $\pm4$ kg m$^{-3}$ levels. Data from run B of \citet{Navarrete22}.
}
\label{fig:PB_rho_contour}%
\end{figure}

In stellar dynamo simulations, non-axisymmetric magnetic fields are
commonly found and tend to be stronger than the axisymmetric modes
at sufficiently rapid rotation \citep{Viviani18}.
The magnetic pressure ($P_\mathrm{B}$) and non-axisymmetric density field
($\rho^\prime$) of run B of \cite{Navarrete22} are shown in Fig.~\ref{fig:PB_rho}
for illustrative purposes,
where the top (bottom) panels are cuts at a latitude of $\theta=+60^\circ$
($\theta=-60^\circ$). These two quantities are  defined as
\begin{equation}
    P_{B} = \frac{B_{\rm rms}^2}{2\mu_{0}},
\end{equation}
and
\begin{equation}
    \rho^\prime = \rho - \langle\rho\rangle_{\phi},
\end{equation}
respectively, where $B_{\rm rms} = \sqrt{(B_{r}^2 + B_{\theta}^2 + B_{\phi}^2)}$
is the root mean square magnetic field and $B_{r,\theta,\phi}$ are the
components of the magnetic field, and $\mu_{0} = 4\pi\times10^{-7}$~H~m$^{-1}$
is the vacuum permeability. Furthermore, $\rho$ is the density and
$\langle\rho\rangle_{\phi}$ is the azimuthally averaged (mean) density field.
In this run the magnetic pressure follows a non-axisymmetric
distribution and the density shows a clear non-axisymmetric
signal. The non-axisymmetric
density field has an extremum of $\pm 4.5$~kg~m$^{-3}$ that closely follows
the magnetic pressure. Upon closer inspection and as depicted in
Fig.~\ref{fig:PB_rho_contour} where $\rho^\prime$ is shown as contours
on top of $P_\mathrm{B}$, the largest values of $\rho^\prime$ lie near the
regions where $P_\mathrm{B}$ is the largest.
This shows that non-axisymmetric density fluctuations are expected
when a non-axisymmetric magnetic field is present as required by the model
of \citet{Lanza20}. However, it should be noted that Figures~\ref{fig:PB_rho}
and \ref{fig:PB_rho_contour} are snapshots and the dynamical evolution of
the magnetic field complicates the picture. In fact, the presence of ADWs moves these fields from the perspective of the rotating frame of
reference of the star, and small-scale non-axisymmetric structures are destroyed and created as the dynamo action progresses.

\begin{figure}
\centering
\includegraphics[width=\columnwidth]{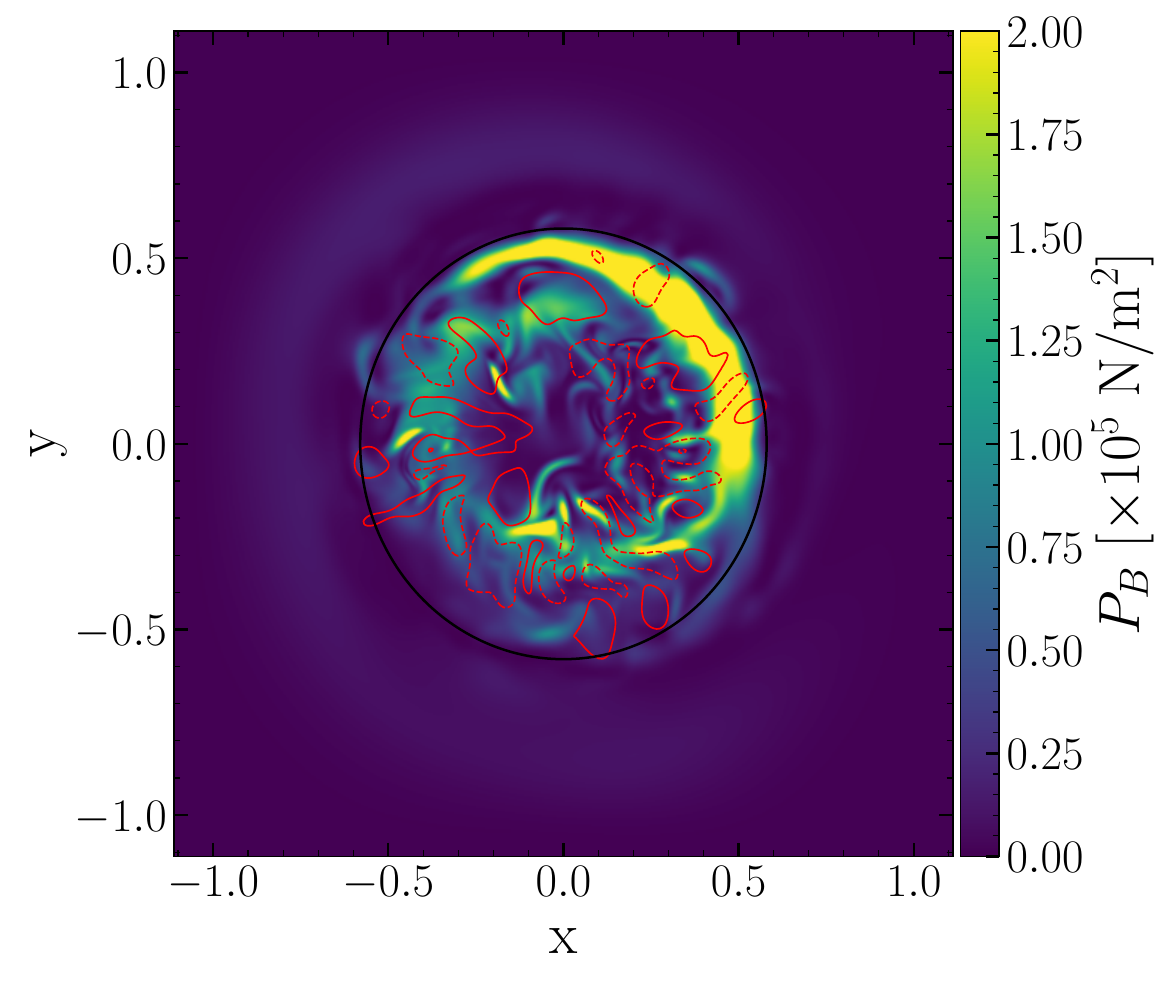}
\caption{Magnetic pressure of an MHD model. The red isocontours correspond
to density variations at the $\pm 5250$~kg~m$^{-3}$ level, and the stellar
surface if shown in the black line.}
\label{fig:PB_rho_Mdwarf}%
\end{figure}

A similar exercise can be done for the star-in-a-box setup
\citep{Kapyla20a}, where a star is embedded in a Cartesian box.
In Fig.~\ref{fig:PB_rho_Mdwarf} we show the magnetic pressure
of an M~dwarf rotating at 30 times the solar rotation rate together
with the isocontours of density variations in red at a height of
$z=0.81$. The black line corresponds to the surface of the star at
this height.
In this case, the relation between $P_B$ and $\rho'$ is more difficult to
see by eye, but there is still a degree of asymmetry in both of them.

\subsection{Azimuthal dynamo waves}\label{sec:ADWs}

ADWs are waves that rotate rigidly in the azimuthal direction in the
frame of reference of the star. They are unaffected by fluid advection 
and are thus decoupled from differential rotation. They can have either a
prograde or retrograde propagation. Despite having been predicted by \citet{Krause80}
with dynamo theory arguments, ADWs have not been studied in detail yet.
An example of an ADW is the westward drift of Earth's magnetic field.

From theoretical grounds, non-axisymmetric fields are expected to be excited
when the $\Omega$ effect is small compared to the $\alpha$ effect.
This is expected for rapidly rotating stars where differential rotation
is quenched and $\alpha$ effect enhanced.
In other words, stars where an $\alpha^2$ dynamo operates should
have stronger non-axisymmetric magnetic fields.
Hydrodynamic and magnetohydrodynamic convection simulations have shown
that the amplitude of differential rotation decreases with rotation
\citep[e.g.][]{Gastine14,Viviani18,Kapyla_et_al_2023_SSRv_219_58}.
As rotation increases, the amplitude of the magnetic field strength increases as well and tends
to produce mostly non-axisymmetric fields, thus facilitating the apparition of ADWs of low azimuthal degree ($m=1,2$) that dominate the simulations
\citep{Cole14,Viviani18}. This is further exacerbated by explicitly including the 
centrifugal force in the momentum equation \citep{Navarrete23}.
More recently, ADWs have also been produced in rapidly rotating fully convective
stars in a setup in which the star is embedded in a Cartesian box
\citep[know as the star-in-a-box setup; see][]{Kapyla20a}.
Generally speaking, magnetic fields introduce density perturbations in an
otherwise uniform medium because magnetic pressure contributes to the pressure balance.
Such density perturbations are directly related to the magnetic field and a
non-axisymmetric field would produce non-axisymmetric perturbations.

The connection between ADWs and ETVs in close binaries is motivated by the
study of \citet{Lanza20} and the evidence of ADWs from
simulations. The assumptions made by \citet{Lanza20},
namely asynchronisation between orbital and rotational period, and a
permanent flux tube, were invoked in order to introduce a non-axisymmetric quadrupole
term in the gravitational potential due to the flux tube, and a periodicity due to asynchronous rotation (see Sect.~\ref{sec:applegate-lanza}).
We here propose that, by having ADWs in the magnetically active component of close binaries, we achieve the same physical mechanism as that of \citet{Lanza20}. Indeed, tidal locking implies that these stars
rotate from fifty times the solar rotation rate
($\Omega_{\odot}$) for V471~Tau, to as high as a few hundred times
$\Omega_{\odot}$ for systems such as NN~Ser and NY~Vir. An $\alpha^2$ dynamo should be present in these stars, and strong
ADWs are expected to be in the main-sequence components of PCEBs.

\section{Model}\label{sec:model}

We solve the dynamics of a binary star with a further contribution that comes
from the quadrupole moment of the secondary. In general, the quadrupole part of
the gravitational potential in Eq.~\ref{eq:gravpot} can be explicitly written as
\begin{equation}
    \phi_{\rm quad} = -\frac{3G}{2r^5}\left(Q_{xx}x^2 + Q_{yy}y^2 + Q_{zz}z^2
    + Q_{xy}xy + Q_{xz}xz + Q_{yz}yz\right).
\end{equation}
The forces associated to $\phi_{\rm quad}$ are \citep{Murray99}
\begin{align}
    F_x & = -m \frac{\partial \phi}{\partial x}\hat{x},\\
    F_y & = -m \frac{\partial \phi}{\partial y}\hat{y}.
\end{align}
The off-diagonal terms are neglected because they are orders of magnitude
smaller than the diagonal terms. The accelerations acting on the $x-$ and
$y-$directions are
\begin{align}
    a_x & = -\frac{3G}{2r^5}\left[Q_{xx}\left(2x-5 \frac{x^3}{r^2}\right)
    - 5Q_{yy} \frac{xy^2}{r^2} - 5Q_{zz}\frac{xz^2}{r^2}\right]\hat{x}, \\
    a_y & = -\frac{3G}{2r^5}\left[-5Q_{xx}\frac{x^2y}{r^2}
+ Q_{yy} \left(2y - 5\frac{y^3}{r^2}\right) - 5Q_{zz}\frac{yz^2}{r^2}\right]\hat{y}.
\end{align}
We can further assume that the orbital motion is constrained to the $x-y$ plane
which coincides with the equatorial plane of the magnetically active star,
assuming that the rotational axis is perpendicular to the orbital plane. This
implies that $z=0$ and thus the accelerations are
\begin{align}
    a_x & = -\frac{3G}{2r^5}\left[Q_{xx}\left(2x-5 \frac{x^3}{r^2}\right)
    - 5Q_{yy} \frac{xy^2}{r^2}\right]\hat{x}, \\
    a_y & = -\frac{3G}{2r^5}\left[-5Q_{xx}\frac{x^2y}{r^2}
+ Q_{yy} \left(2y - 5\frac{y^3}{r^2}\right)\right]\hat{y}.
\end{align}
We implement these accelerations in the N-body solver {\sc Rebound}
\citep{rebound} with the \texttt{additional\_force} method. We use
a 15th-order Gauss-Radau integrator (IAS15) to solve the motion of the binary
\citep{reboundias15}.

\begin{figure*}
\centering
\includegraphics[width=\textwidth]{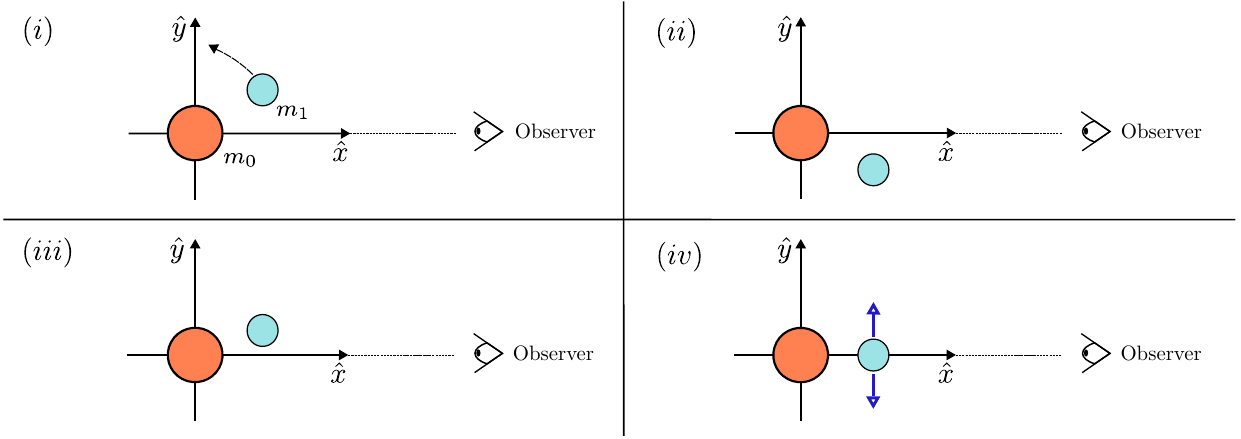}
\caption{Simple schematics of the transit times algorithm. ($i$): the simulation
is initialized with $m_{0}$ as the mass of the white dwarf and $m_{1}$ as the mass
of the magnetically active star. The dashed line indicates the direction of the
orbital motion. The observer is placed at $(+\infty,0)$. $(ii)$: after a few
integrations, the $m_{1}$ star is right under $y=0$ and $x>0$. The simulation
time is stored. $(iii)$: one more integration is performed, and the new
$y$-position ($y_{\rm new}$) is compared to previous one $y_{\rm old}$.
A transit must have happened between the current and previous times because
$y_{\rm new}$ > $y_{\rm old}$. $(iv)$: the simulation is integrated backwards
with a timestep equal to half the difference between the current and previous
times. A new integration forward in time takes place. This is done iteratively,
as indicated by the blue line, until the difference between the current and previous
time is less than $10^6$ times smaller than an orbital period. The time when
this condition is fulfilled is stored as a transit time and the process continues from ($i$).
}
\label{fig:transits}%
\end{figure*}

\subsection{Simulation strategy}\label{sec:strategy}
Each set of binary parameters, namely the masses of the stars and the distance
between the two, corresponds to one set of simulations. Each simulation is
controlled by a total of five free parameters. These are the values
of $Q_{xx}$ and $Q_{yy}$, their periods which we take to be the same, consistent with results from \citet{Navarrete22} and consistent with the idea that both have the same origin, the mass of the white dwarf,
and the binary separation. The mass of the magnetically active star that hosts the
non-axisymmetric quadrupole moment is fixed at $0.2\msol$. The quantities related
to the quadrupole moment are uncertain as they
have not been derived from observations. However, we can nonetheless employ results obtained through simulations of stellar dynamos as a guide. 

Each simulation is started assuming a circular orbit, and the initial
velocities are set accordingly. The two components of
the quadrupole moment, namely $Q_{xx}$ and $Q_{yy}$, are re-computed
on each timestep according to the prescribed migration period of the ADW.

To search for transits, we place an observer along the $x$-axis at infinity and
at $y=0$ in a coordinate system originating at the magnetically
star active star (see Fig.~\ref{fig:transits}). This observer
would see a transit when the position of the $m_1$ star changes from
$y_1<0$ to $y_1>0$ and when $x_1>0$, where the subscript `1' refers to the
companion star. In practice, we integrate the binary orbits and check for
this condition. When a change of sign in $y_1$ is detected, we integrate
backwards between the current and last timestep by bisection until the time
difference between the current and last timesteps is small enough.
This difference is, in general, about $10^6$ times smaller than the time it
takes to complete an orbit. These transit times will have a linear contribution
coming from the time it takes to complete an orbit. We remove this trend with
a linear least square fit, thus leaving us with just the
transit time variations.\footnote{This method is based on the
`TransitTimingVariations' IPython example in {\sc Rebound}'s GitHub repository.}

\subsection{MHD models and input quadrupole moment}\label{sec:input}
{
In the present simulations we are only concerned about the effects of
ADWs that have rotation but we neglect possible  modulation effects. That
is, the amplitude of the ADWs does not change with time.
For this purpose, we work with snapshots of three mature MHD models using the setup of \citet{Kapyla20a, Ortiz2023} of
fully convective $0.2\msol$ stars. We name these MHD simulations as Model A with
a stellar rotation rate of $10~\Omega_\odot$, Model B with $20~\Omega_\odot$,
and Model C $30~\Omega_\odot$.
The resolution of the MHD models are $288^3$ for the Model A,
and $576^3$ for models B and C. These MHD runs are characterized
by dimensionless system parameters. These are the stellar Coriolis
number (${\rm Co}_\star$), Taylor number (${\rm Ta}$), the sub-grid
scale (SGS)
and magnetic Prandtl numbers (${\rm Pr}_{\rm SGS}$ and ${\rm Pm}$) and
which are defined as
\begin{align*}
    {\rm Co}_\star &= 2 \Omega_0 \left(\frac{3MR^2}{L} \right)^{1/3}, \hspace{0.1em}
    & {\rm Ta} &= \frac{4\Omega_0R^4}{\nu^2}, \\
    {\rm Pr}_{\rm SGS} &= \frac{\nu}{\chi_{\rm SGS}}, \hspace{0.1em}
    & {\rm Pm} &= \frac{\nu}{\eta},
\end{align*}
where $\Omega_0$ is the rotation rate, $M$, $R$, and $L$ are the
stellar mass, radius, and luminosity, respectively, $\nu$ is the
kinemetic viscosity, $\chi_{\rm SGS}$ is the SGS entropy diffusion, and $\eta$
is the magnetic diffusivity. ${\rm Co}_\star$ is related to the
modified diffusion-free flux-based Rayleigh number (${\rm Ra}_{\rm
  F}^\star$) via ${\rm Co}_\star = ({\rm Ra}_{\rm F}^\star)^{-1/3}$
\citep[e.g.][]{Kapyla_2023_AA_669_98, Kapyla_2024_AA_683_221}. We also
give the fluid and magnetic Reynolds number (Re and Re$_M$) defined as
\begin{equation*}
    {\rm Re} = \frac{u_{\rm rms} R}{2\pi\nu},\,\,\, {\rm Re}_M = \frac{u_{\rm rms} R}{2\pi\eta},
\end{equation*}
where $u_{\rm rms}$ is the root-mean-square fluid velocity.
These values are summarized in Table~\ref{tab:input_mhd}.  For further
details on the model configuration, see \citep{Kapyla20a}.
\begin{table}
  \caption{Dimensionless system parameters and diagnostics for the MHD models.}
  \label{tab:input_mhd}
  \begin{center}
    \begin{tabular}{c|c|c|c|c|c|c}
      \hline
      Model & ${\rm Co}_\star$ & Ta & ${\rm Pr}_{\rm SGS}$ & Pm & Re & Re$_M$\\
      \hline
      \hline
      A & 768  & $1.60\times10^{11}$ & 0.2 & 0.5 & 73  & 36 \\
      B & 1536 & $6.40\times10^{11}$ & 0.2 & 0.5 & 49  & 25 \\
      C & 2304 & $5.76\times10^{12}$ & 0.2 & 0.5 & 100 & 50 \\
      \hline
    \end{tabular}
  \end{center}
\end{table}

Each model is characterized by the input quadrupole moment taken from 
direct numerical simulations of stellar convection with magnetic fields. We 
select snapshots that correspond to a maximum of magnetic energy. The 
snapshots are then rotated around a fixed coordinate system to calculate
the effective quadrupole moment the binary system
would feel as a function of the phase of the system $\phi$, which is then used as 
input for the calculations. The obtained quadrupole moments as a function of 
phase are provided in Appendix~\ref{app:input}. When applying this
to the simulations of the binary systems, it allows us to map it onto different
time evolutions by assuming different possible periods of the ADWs.

We note that the centrifugal force is turned off in all of the
MHD models, which is a common approximation for stellar dynamo
simulations. In the context of objects close to spherical
symmetry, this could usually be considered as a good
approximation. There are in addition numerical reasons previously
outlined by \citet{Kapyla20b}, which require to scale down the
strength of the centrifugal force in fully compressible simulations
where the enhanced luminosity
approximation is employed. We have investigated the implications of
including the centrifugal force in a previous study, where we found
that its impact on the quadrupole moment is not significant, even
though the dynamo solution is being affected due to the
redistribution of magnetic energy from the axisymmetric mode to the
non-axisymmetric mode \citep{Navarrete23}. As here we are only
interested in obtaining the magnitude of the quadrupole moment from
the simulations, we consider it as a safe approach to neglect the
centrifugal force.

We first construct synthetic quadrupole moments from moments of inertia assuming that they follow
a sinusoidal shape, that is
\begin{equation}
  I_{ii} = I_{ii,\,{\rm m}}^{\rm ref} + f \Delta I_{ii}^{\rm ref} \sin\left(\frac{t}{t_{\rm mod}}\right),
\end{equation}
where $ii = {xx, yy, zz}$, $I_{ii,\,{\rm m}}^{\rm ref}$ are the mean values of
the moment of inertia of the reference run, and 
$\Delta I_{ii}^{\rm ref}$ the amplitude of their variations. We also include a factor $f$ which
parametrizes possible modifications in the amplitude of the variations. Each set of  
simulations is named after the value of $f$, see Table~\ref{tab:sinruns}.
From these moments of inertia, we compute the three diagonal components of $Q$ as
\begin{align}
Q_{xx} &= I_{xx} - \left(I_{xx} + I{yy} + I_{zz}\right)/3, \\
Q_{yy} &= I_{yy} - \left(I_{xx} + I{yy} + I_{zz}\right)/3, \\
Q_{zz} &= I_{zz} - \left(I_{xx} + I{yy} + I_{zz}\right)/3, 
\end{align}
which enter in the computation of the accelerations in our model.
The reference values of the moment of inertia are taken from Model A,
which are summarized in Table~\ref{tab:input_intertia}.
}

\begin{table}
  \caption{Summary of simulations with sine-wave moment of inertia components.}
  \label{tab:sinruns}
  \begin{center}
    \begin{tabular}{c | c | c | c | c | c | c}
    \hline
       & sin1p0 & sin1p2 & sin1p4 & sin1p6 & sin1p8 & sin2p0 \\
    \hline
    \hline
      $f$ & 1.0 & 1.2 & 1.4 & 1.6 & 1.8 & 2.0 \\
    \hline
    \end{tabular}
    \tablefoot{For each set, we vary the mass of the secondary star
    $m_1 = 0.80,\, 0.67,\, 0.53,\, 0.40\, \msol$, the binary separation
    $a_{\rm bin} = 1.0,\, 0.75,\, 0.5\, \rsol$, and eight equally-spaced values of
    $P_{\rm mod}$ between 20 and 8 years, resulting in a total of 96 simulations
    per set.}
  \end{center}
\end{table}

\begin{table*}
  \caption{Reference values of the inertia tensor used for the sine wave models.
  All values are in units of kg m$^{2}$ and parenthesis indicate exponents of 10.}
  \label{tab:input_intertia}
  \begin{center}
    \begin{tabular}{c|c|c|c|c|c}
      \hline
      \hline
      $I_{xx,{\rm m}}^{\rm ref}$ & $I_{yy,{\rm m}}^{\rm ref}$ & $I_{zz,{\rm m}}^{\rm ref}$ & $\Delta I_{xx,{\rm m}}^{\rm ref}$ & $\Delta I_{yy,{\rm m}}^{\rm ref}$ & $\Delta I_{zz,{\rm m}}^{\rm ref}$ \\
      \hline
      4.44172442(45) & 4.44172441(45) & 4.39161589(45) & 6.00192126(40) &
      6.00267489(40) & 1.40925322(41) \\
      \hline
    \end{tabular}
  \end{center}
\end{table*}

\section{Results}\label{sec:results}

To validate our model and to better understand
the system and the expected \omc\, amplitudes and periods, we study the results
from moment of inertia fluctuations  modeling them as sine waves as described
above and constructing the corresponding \omc\, diagrams.  In what follows,
we use $P_{\rm mod}$ and $P_{\rm ADW}$ interchangeably, as we assume that the modulation period is due to the ADWs.
We vary all of the parameters that affect the binary motion, except
the secondary's mass. This is because the MHD simulations from which we take
the reference values of the moment of inertia are fixed at $0.2\msol$
as described in Sect.~\ref{sec:input}.


\begin{figure}
\centering
\includegraphics[width=\columnwidth]{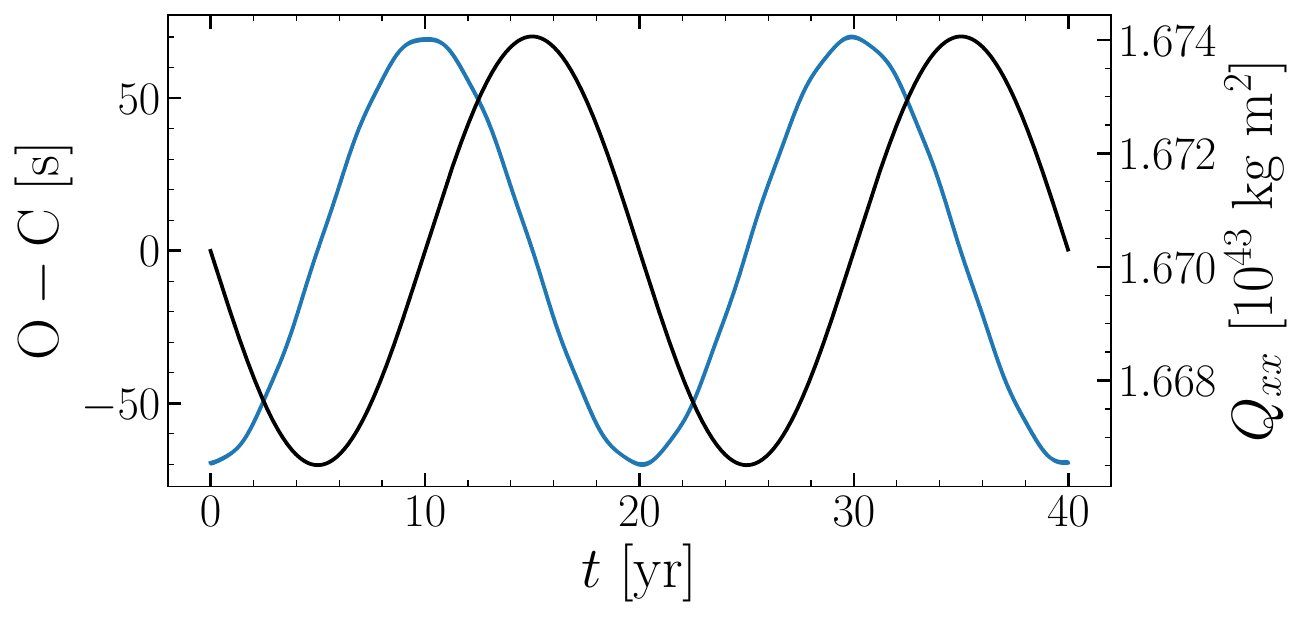}
\caption{\omc\, diagram (blue) for the eclipsing times of a run {corresponding to}
$m_{0}=0.2\msol$ and $m_{1}=0.8\msol$, binary separation
$a_{{\rm bin}}=1\rsol$, and $P_{{\rm ADW}}~=~20$~yr. The quadrupole
moment $Q_{xx}$ is shown in black.}
\label{fig:omc}
\end{figure}

\begin{figure*}
    \centering    \includegraphics[width=\textwidth]{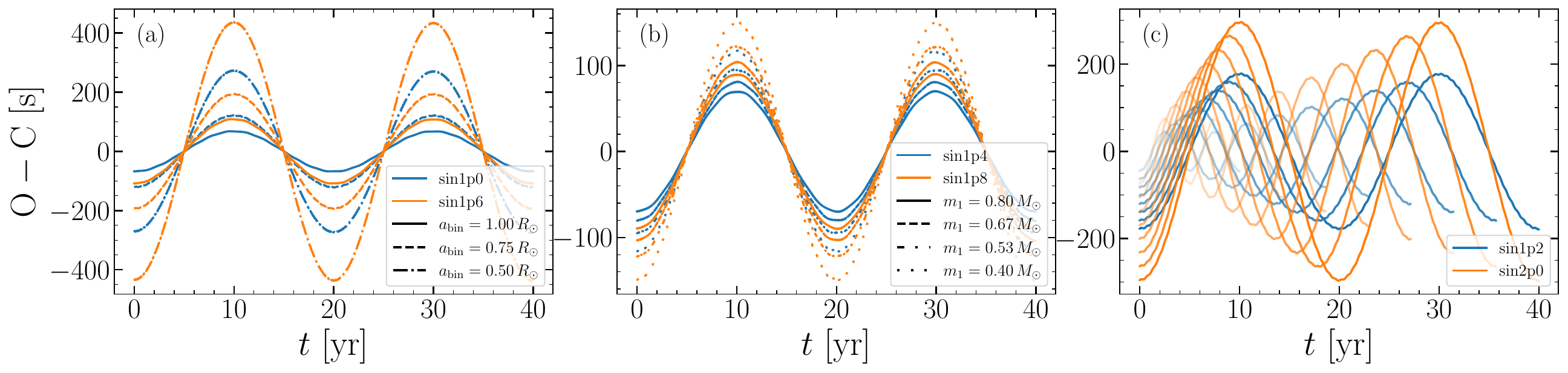}
    \caption{\omc\, diagrams for different runs: (a) For sets sin1p0 with $f=1.0$ and sin1p6 with $f=1.6$
      for $P_{\rm mod}=20$ yr and
      $m_1 = 0.53\msol$ with varying binary separations $a_{\rm bin}$ ranging from $0.5\rsol$ to $1.00\rsol$. (b) For sets sin1p4 with $f=1.4$ and sin1p8 with $f=1.8$ where
      $a_{\rm bin} = 1.0\rsol$ and $P_{\rm mod} = 20$ yr varying $m_1$ between $0.40\msol$ and $0.8\msol$. (c) For sets
      sin1p2 with $f=1.2$ and sin2p0 with $f=2.0$ where
      $a_{\rm bin} = 0.75\rsol$, $m_1=0.4\msol$, while $P_{\rm mod}$ varies between $5$~yr and $20$~yr, with fainter lines corresponding to smaller $P_{\rm mod}$.
    }
    \label{fig:omc_diag}
\end{figure*}

A typical \omc\, diagram is shown in Fig.~\ref{fig:omc}. 
It reaches an amplitude of about 50~s, even in the case where the quadrupole 
moment amplitude is rather small. The period is 20~yr as expected from the period of
the quadrupole moment, and is repeated twice, which coincides with the 
quadrupole moment. We find a phase shift between the \omc\, and $Q_{xx}$ 
signals of about five years. As the \omc\, diagram corresponds to an integral 
over the period changes $\Delta P$, it implies that the \omc\, diagram will 
scale as $-\cos( t/t_{\rm mod} )$ if
$\Delta P/P\propto Q_{xx}\propto\sin( t/t_{\rm mod} )$. It implies a phase 
shift by $\pi/2$, corresponding to a quarter of a cycle.

We now explore how the \omc\, diagrams change when the binary parameters are
varied. For illustrative purposes, we take sets sin1p0 and sin1p4 and plot the
\omc\, diagrams as a function of time for different values of $m_1$ and $a_{\rm bin}$
while the other parameters are kept constant. In panel (a) of Fig.~\ref{fig:omc_diag}, we show how the \omc\, diagram changes when the
binary separation is changed. For sin1p0, the minimum \omc\, amplitude is
$68$~s and increases to $274$~s when the binary separation decreases to
$0.5\rsol$. For Set sin1p6 with $f=1.6$, these numbers are $108~\mathrm{s}$ and
$437~\mathrm{s}$, respectively. The scaling of the \omc\, amplitude with binary
separation can be understood by comparing Eq.~\ref{eq:Delp_Qxx} with
\begin{equation}
  \frac{\Delta P}{P} = 2\pi\frac{\omcl}{P_{\rm mod}},
\end{equation}
which gives
\begin{equation}\label{eq:scaling_omc}
  \omcl = \frac{9}{2\pi}\frac{\Delta Q_{xx}}{a^2m_0}P_{\rm mod}.
\end{equation}
Comparing two simulations with different binary separation but otherwise
identical parameters yields
\begin{equation}
  \frac{\omcl_1}{\omcl_2} = \frac{a_2^2}{a_1^2},
\end{equation}
where the indices refer to different simulations. We checked that this relation holds for all of
our simulations, which is an indication that our model behaves as expected.

Another important binary parameter is the mass of the companion star. This dependence is shown in panel (b) of
Fig.~\ref{fig:omc_diag}. The peak of \omc\, is achieved for  the smallest companion mass
$m_1$  for both sets, and it is reduced as $m_1$ increases. While the
mass is increased by a factor of two, from $0.4\msol$ to $0.8\msol$, the
inverse  ratio of the peaks of the \omc\, is 1.67 for both sets.

Finally, we show how the \omc\, diagram changes when only $P_{\rm mod}$ is changed.
This is shown in panel (c) of Fig.~\ref{fig:omc_diag}. The amplitude of \omc\,
grows linearly with $P_{\rm mod}$, reaching a peak of 178 and 296 seconds for
sets sin1p2 and sin2p0, respectively, when $P_{\rm mod} = 20$~yr. This scaling
is also understood from Eq.~\ref{eq:scaling_omc}, which in this case gives
\begin{equation}
  \frac{\omcl_1}{\omcl_2} = \frac{P_{\rm mod, 1}}{P_{\rm mod, 2}}.
\end{equation}
That is, for longer migration periods of the ADWs with respect to the rotating frame of the
magnetically active star, the larger the amplitude in the \omc\, diagram.

\begin{figure}
    \centering
\includegraphics[width=\columnwidth]{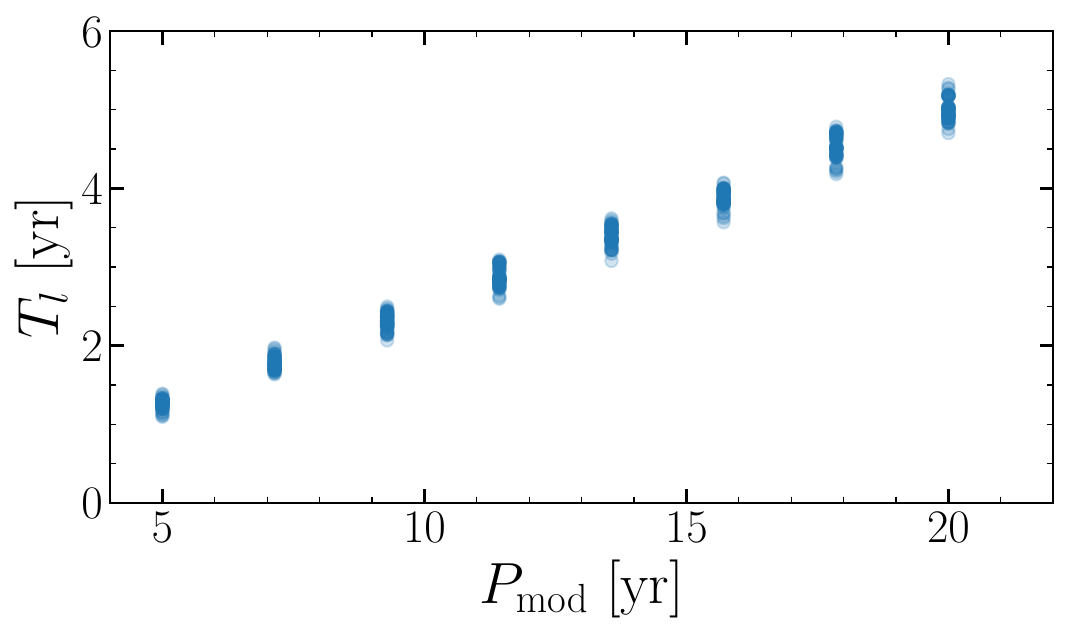}
    \caption{Time lag $T_l$ as defined in Eq.~\ref{eq:Tl} between $\Delta Q_{xx}$ and the \omc\, diagram for
      all runs shown as a function of the modulation period of the variations.
    }
    \label{fig:Tl}
\end{figure}

As a consistency check for our interpretation of the \omc\, diagram, we also check if the time lag between the change of the quadrupole moment and the \omc\, diagram behaves as expected. We define the quantity
\begin{equation}\label{eq:Tl}
  T_l = t_{Q}^{\rm peak} - t_{\omcl}^{\rm peak},
\end{equation}
which we plot in Fig.~\ref{fig:Tl} as a function of the modulation period of
the dynamo wave, where each dot represents a different simulation. The time lag scales linearly with $P_{\rm mod}$ and corresponds to about a quarter of the modulation period. The scatter in $T_l$ is due to the different quadrupole moment amplitudes and secondary masses used
in our simulations. The origin of the time lag can be understood from the fact that the
\omc\, diagram corresponds to a time integral over the period variations.
If $\Delta P / P \propto Q_{xx}(t)$ and we approximate
$Q_{xx}(t) \propto \sin(2\pi t / P_{\rm mod})$, then
\begin{equation}
\omcl \propto \int \frac{\Delta P}{P} \, dt 
\propto \int \sin\!\left(\frac{2\pi t}{P_{\rm mod}}\right) dt
\propto -\cos\!\left(\frac{2\pi t}{P_{\rm mod}}\right),
\end{equation}
which implies a phase shift of $\pi/2$, corresponding to a time lag
$T_l \simeq P_{\rm mod}/4$. 

The fact that $T_l$ is non-zero has observational
consequences because correlations between the magnetic field state of a star in
a PCEB and the eclipsing time variations will have a corresponding phase shift as well. This means that
attempts to explain ETVs with magnetic activity need to consider this time lag. 

We also perform simulations where the time-dependent
quadrupole moment is not based on an analytical function, but taken from the MHD
simulations described above (models A, B, C) that have been rotated for different
angles to account for different possible phases of the system. As a starting
point, we consider Model A with 10 times solar rotation for a $0.2\msol$
star. 

\begin{figure}
    \centering
\includegraphics[width=\columnwidth]{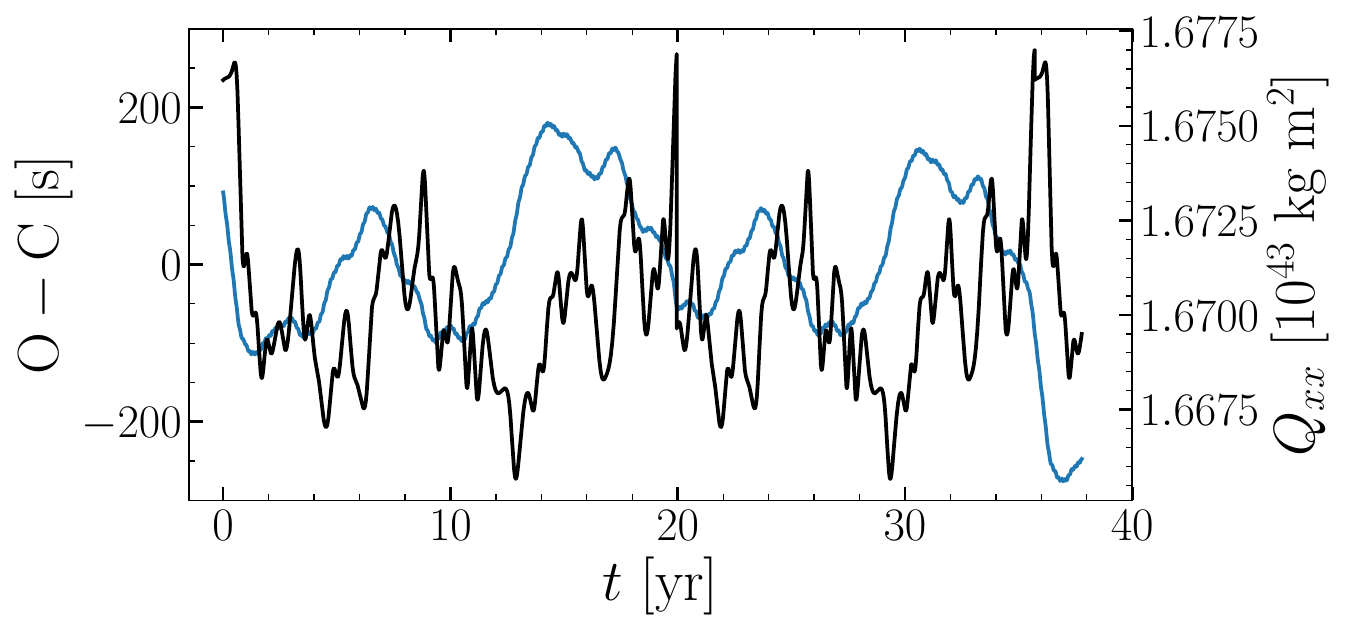}
    \caption{\omc\, diagram (blue) for the eclipsing times of a run with a
      quadrupole moment directly taken from an MHD simulation {of a
      fully-convective star { with a rotation rate of $10\Omega_\odot$
      (Model A)}}. The quadrupole
      moment $Q_{xx}$ is shown in black. {A detailed description of the
      MHD setup can be found in \citet{Kapyla20a}}.}
    \label{fig:omc_new}
\end{figure}

To emulate the effects of a migrating non-axisymmetric magnetic field,
we work by taking a single snapshot of an MHD simulation. Choosing one single
snapshot instead of a series of them allow us to focus on studying purely the
effects of rotating ADWs as unaffected by the modulation of the amplitude which
would otherwise complicate the interpretation of the results. We rotate the
density field of the MHD simulations about the rotational axis of the star.
The components of the quadrupole moment are computed each time about a non-rotated
coordinate system.
These values are stored and used as input to our N-body simulations.
The rotation angle can be translated to an arbitrary time, which represents the
period of migration of the ADW, i.e., $\phi = \pi$ corresponds to half of the
period of the ADW. The time-dependent quadrupole
moment and the resulting \omc\, diagram is show in Fig.~\ref{fig:omc_new}.

Similarly to the previous results, the \omc\, diagram lags behind the quadrupole moment. This is more difficult to see here because of the more complex
time evolution of the quadrupole moment, but there are still some features than can be distinguished by
eye. For example, when $\omcl = 0$ is crossed, $Q_{xx}$ reaches a maximum
or a minimum, and the maximum value of $Q_{xx}$ is ahead of the maximum of
\omc\, by $3\dots5$ years.
The small, short-period changes in the quadrupole moment are not reflected in
the \omc\, diagram as small variations are washed out due to the time integration over the period changes. A rather complex time evolution of the quadrupole moment can thus mimic a superposition of quasi sinusoidal modes in the \omc\, diagram for time spans corresponding to a few decades. 

We perform 96 N-body simulations with varying $P_{\rm ADW}$, $a_{\rm bin}$ 
and $m_1$ for each MHD model (models A, B, C) and select the resulting
\omc\, diagrams for the values of $m_1=0.53\msol$, 
$a_{\rm bin} = 0.75\rsol$, and two for $P_{\rm ADW}$, which are shown in
Fig.~\ref{fig:omc_pencil}. Further details on the input quadrupole moments
are in Appendinx~\ref{app:input}.
\begin{figure*}
\centering   \includegraphics[width=\textwidth]{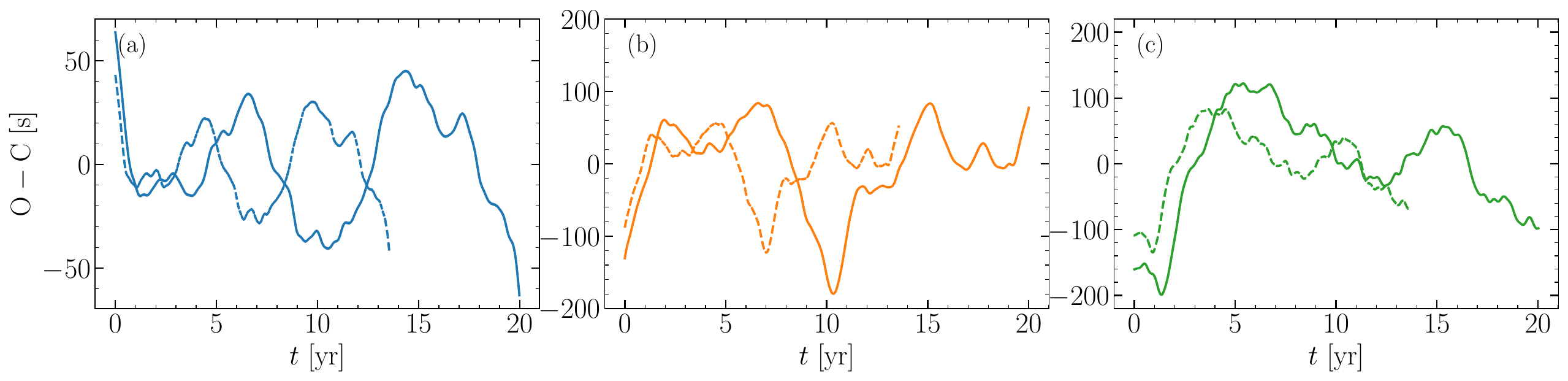}
\caption{\omc\, diagrams for runs that directly used the output of MHD
  simulations,  which differ by increasing stellar rotation rate
  { from $10\Omega_\odot$ (Model A), same simulation as in
  Fig.~\ref{fig:omc_new} in the left panel,
  $20\Omega_\odot$ (Model B) in the middle panel, to $30\Omega_\odot$
  (Model C) in the right panel}.
  Each panel corresponds to a different quadrupole moment,
 and each line to a different $P_{\rm ADW}$, { about 14 years for the
 dashed lines and 20 years for the solid lines}. 
 For each run, $m_1=0.53\msol$ and $a_{\rm bin} = 0.75\rsol$.
    }
    \label{fig:omc_pencil}
\end{figure*}
For ease of comparison, we fix the mass of the companion star and the binary
separation to $m_1=0.53\msol$ and $a_{\rm bin} = 0.75\rsol$.
Depending on the behavior of the quadrupole moment as the density field rotates
and how large it is, different \omc\, diagrams are obtained.

In panel (a) of Fig.~\ref{fig:omc_pencil}, the maximum achieved \omc\, amplitude
 is about $50$ s and the minimum is $30$ s. In this case, the \omc\, is
quasi-periodic with some small-amplitude scatter. There are some episodes where
the \omc\, quickly changes from positive values to negative values within a
five-years period. Following that, a minimum or maximum is attained for just a few
years. The results shown in panel (b) are similar to those of panel (a), but the 
amplitude is larger by about a factor of four around $t=10$ yr for the solid
line ($t=7$ for the dashed line). There are also episodes where the \omc\,
quickly drops and rises by about 300 s. In panel (c), the solution is more stable with fewer drastic drops with the
exception of that in the beginning, which takes the \omc\, from $-200$ s
to $100$ s in six (four) years for the solid (dashed) line. Following this,
there is a steady decrease in the \omc\, that takes place during the rest of
the simulated time. 

\section{Discussion and conclusions}\label{sec:discuss}

We have built an N-body framework aimed to reproduce \omc\, diagrams of
close binaries under the influence of magnetically-induced quadrupole moment
changes. This is motivated by two seemingly unconnected results. The first are
the results of careful ETVs observations of post-common-envelope binaries.
The quasi-periodic behavior of the \omc\, of their eclipsing times is almost
universal as they are present in about 90\% of PCEBs \citep{zorotovic13}.
A second important ingredient in our model is the numerical result that ADWs are easily excited in rapidly rotating
simulations of fully- and partially-convective stars
\citep{Cole14, Viviani18, Kapyla20a, Navarrete23}. Such azimuthally-migrating waves
could explain the mentioned observations extending a previous idea 
presented by \citet{Lanza20}, who explained that presence of a time-dependent quadrupole moment as a result of a fixed quadrupole moment in the reference frame of the magnetically active star together with the assumption of a deviation from tidal synchronization. This assumption can be at least difficult to motivate for compact binary systems, and here as an alternative we consider the quadrupole moment variation to be caused by time-dependent ADWs.

We first tested our model with assuming sinusoidal variations of the quadrupole moment, using typical numbers obtained from MHD simulations to study the reaction of a close binary system to such variations. We were able to obtain \omc\, amplitudes of up to several hundred seconds, with the modulation period being equal to the period  of
the quadrupole moment changes, including a systematic exploration on
the dependence of binary separation, companion mass and modulation
period. We have further shown that there is a phase shift
corresponding to about a quarter of the modulation period between the
quadrupole moment variation and the \omc\, diagram.

We also considered the effect of a more complex quadrupole moment
using input from the dynamo simulation of a $0.2\msol$ star by
\citet{Kapyla20a}. In this case, there is a similar phase shift, but
the evolution of the \omc\, diagram is somewhat more complex and may
roughly resemble a superposition of a few different sinusoidal modes
when considered over times corresponding to a few decades. We note
that the \omc\, diagram is smoother compared to the time-dependent
quadrupole moment as the \omc\, diagram effectively corresponds to a
time-integration over the period variations $\Delta P$.

One of the best studied PCEBs is V471~Tau. Its most up-to-date \omc\, diagram was recently presented by \citet{kundra22} covering now a second cycle. The long-term trend of the \omc\, diagram is about 130 s, with a
period of 35 years, and residuals with a maximum of 45~s and periods of up to
11~years. The \omc\, of V471 Tau is approximately sinusoidal and its
shape  resembles those of Fig.~\ref{fig:omc_diag}, although the amplitude is
smaller in our case but similar to that of Fig.~\ref{fig:omc_new}.
\citet{kundra22} explained the long-term variation by the presence of a
third body, corresponding to a circumbinary brown dwarf of
$0.035\msol$ on an eccentric orbit, and the short-term variations were
attributed to the Applegate or Applegate-Lanza mechanism. While this is
plausible, we find that it is equally possible to explain both the long-
and short-term variations with ADWs only. 

Another well studied PCEB is NN Ser, which has been proposed to host two
planets orbiting the binary \citep{beuermann10}. The proposed planets around
NN Ser were the only ones that correctly predicted future eclipsing times for
several years \citep{marsh14,bours16}. However, in a recent study,
\citet{Ozdonmez23} included more mid-eclipse times of NN Ser and studied the
orbital period variations, and found that the previously proposed model
needed to be revised. While their one-planet solution is stable over
10~Myr, statistically better results are obtained when a second circumbinary 
object is added. However, their two-planets solutions are stable only
in a narrow parameter regime such that only the best-fitting model is stable.
It has been argued that the energy budget of the main-sequence star in NN Ser
is smaller than what is needed to trigger the Applegate mechanism \citep{brinkworth06}
and the Applegate-Lanza mechanism \citep{Lanza20}. However, here we find that
in our model, a $0.2\msol$ star in a close orbit can produce \omc\, variations of hundreds of seconds in the presence of an ADW
in its convective zone.

The case of QS Vir is particularly interesting due to the sharp drop in its
\omc\, diagram between the years 2003 and 2006 \citep[see e.g.][]{bours16}.
Recently, \cite{Giuppone24} has developed a framework to construct \omc\,
diagrams from ephemeris. Focusing on QS Vir, they found that the data is
best-fitted by a third body of $\sim0.055\msol$ on a highly eccentric
orbit and a period of about $16.7$ years. As it is commonly found in PCEBs,
the third-body solution of QS Vir has been found unstable over short periods
of time \citep{Horner13} and thus needed to be revisited. In the study of
\citet{Giuppone24}, some solutions were found to be stable over a span of
$10^6$ yr. However, the authors argue that the residuals should not be modeled
with a fourth body as the solution might become unstable. Thus, in this picture,
the residuals (which can reach values of up to 50 s) could be attributed to
the effects of magnetic fields. We note that also in some of our runs,
as visible for example in Fig.~\ref{fig:omc_new}, relatively abrupt
changes in the \omc\, diagram can occur and it is conceivable that
such a behavior can be explained by intrinsic magnetic processes in
the star. It is particularly important here to consider the complex
internal evolution of the quadrupole moment that could be driven by a
superposition of different dynamo modes sometimes including large
intrinsic fluctuations.

{Most recently, studies by \cite{Pulley2025} and \cite{Yates2026} have called
into question the circumbinary planets solutions as the new data does not fit
the previously proposed models. At the same time, energetic estimates to drive
ETVs via the Applegate or Lanza mechanisms fall below the required levels in all
but a few systems. In this paper, however, we did not test any of those models but
instead tackled the scenario of having a permanent non-axisymmetric magnetic that
rotates in the rotating frame of reference of the star. In such a case, our N-body
simulations result in \omc\, diagrams that are comparable to those that are
obtained from observations.
}

The model proposed here could be further tested through a determination of the magnetic state of the star and particularly its non-axisymmetric component. As surface
magnetic field mapping might not be adequate for rapidly rotating stars
\citep{Zaire22}, we suggest that the improvement of magneto-asteroseismology
techniques could offer such possibilities. This could be complemented by performing
a similar study to what has been presented here, but targeting real systems.
By modeling a specific, observed \omc\, diagram along with its binary parameters, one may aim to determine the time evolution of the quadrupole moment that reproduces the observed diagram, thereby probing the underlying mechanism that affects the density field inside the star. This could allow to infer information about the non-axisymmetric magnetic field.

In summary, our results demonstrate that magnetically induced,
time-dependent quadrupole moment variations driven by ADWs
provide a viable explanation for the quasi-periodic \omc\,
variations observed in many post-common-envelope binaries. Within this
framework, both the amplitudes and the timescales of the observed
signals can be naturally reproduced without additional bodies, while
also accounting for irregular features through the superposition of
multiple dynamo modes. This offers a unified interpretation of systems
such as V471~Tau, NN~Ser, and QS~Vir, where purely dynamical
explanations face stability or fine-tuning challenges. Although
further observational and theoretical work is required to firmly
establish this mechanism, particularly through constraints on stellar
magnetic field geometries, our study highlights the potential of \omc\,
analyses as a probe of internal stellar magnetism, underscoring the
importance of incorporating magnetic processes into the interpretation
of eclipse timing variations in close binaries.

\begin{acknowledgements}
    FHN acknowledges partial support from the program Unidad de
    Excelencia Mar\'ia Maeztu, reference CEX2020-001058-M.
    This research was supported by the Munich Institute for Astro-, Particle and
    BioPhysics (MIAPbP) which is funded by the Deutsche Forschungsgemeinschaft
    (DFG, German Research Foundation) under Germany's Excellence Strategy – EXC-2094 – 390783311.
    The simulations were ran at the NHR@ZIB cluster. DRGS  gratefully acknowledges the support of the ANID BASAL project FB21003 and the Alexander von Humboldt - Foundation, Bonn, Germany. 
\end{acknowledgements}

\bibliographystyle{aa} 
\bibliography{bibliography} 

\begin{appendix}

\section{Input quadrupole moment}\label{app:input}

\begin{figure*}
\centering   \includegraphics[width=\textwidth]{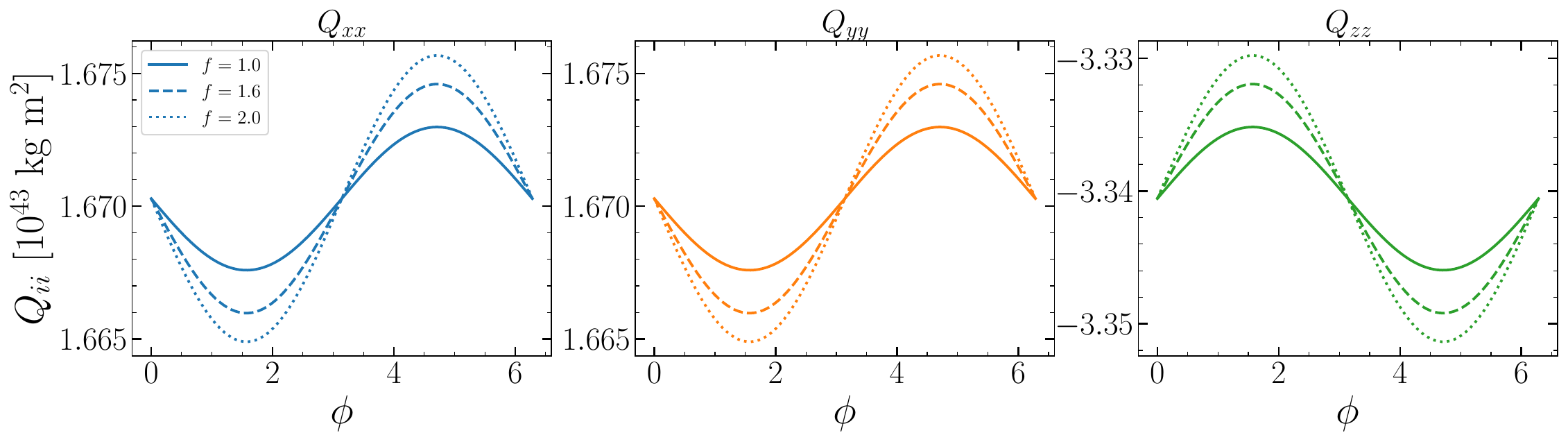}
\caption{Input values of the diagonal values of the quadrupole moment for the
simulations with sinusoidal signals plotted as a function of the phase angle $\Phi$. Each panel shows the corresponding values
for $f=1.0$ (solid line), $f=1.6$ (dashed lined), and $f=2.0$ (dotted line).}
\label{fig:input_sin}
\end{figure*}

\begin{figure*}
\centering   \includegraphics[width=\textwidth]{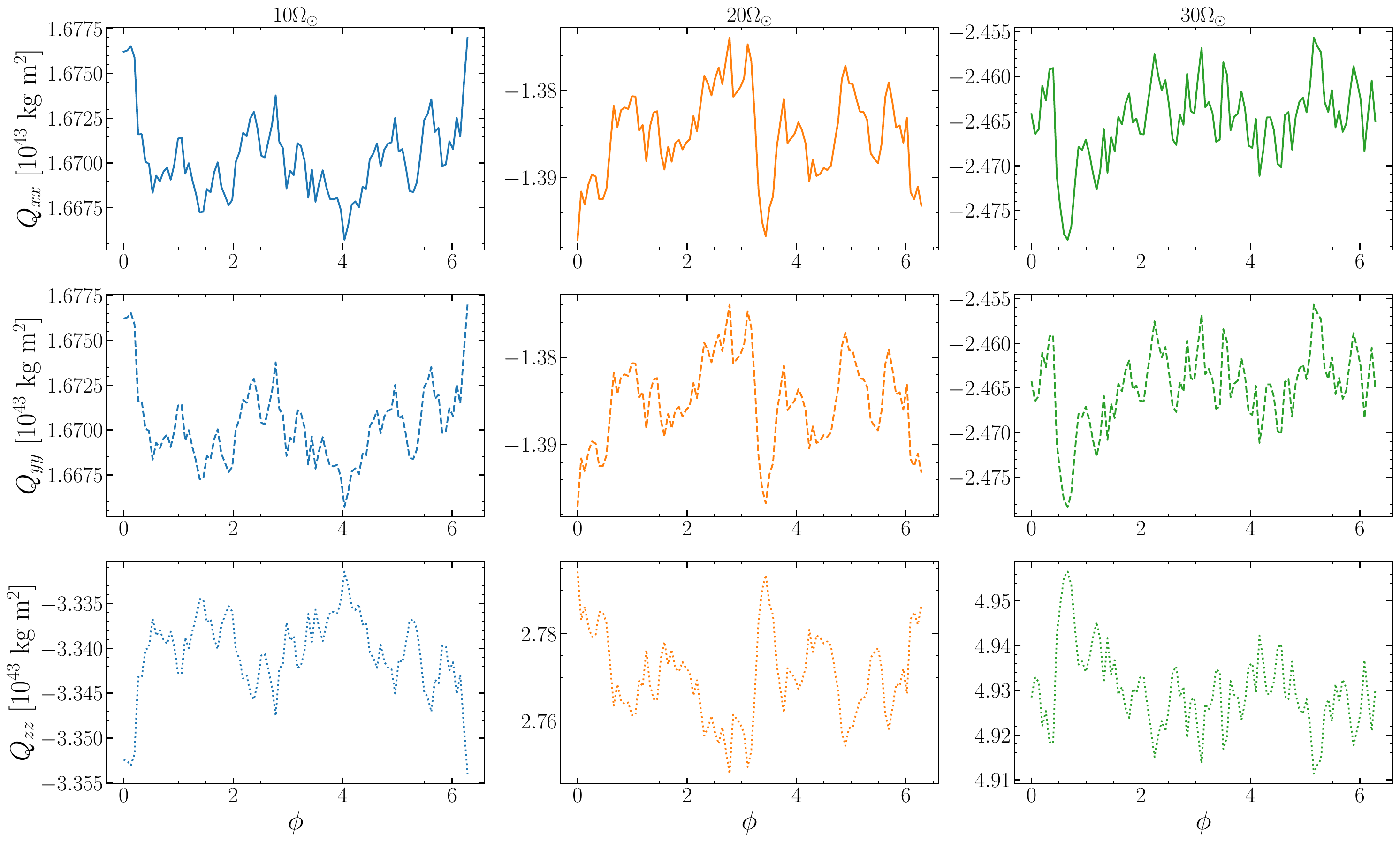}
\caption{Input values of the diagonal values of the quadrupole moment for the
simulations where $Q_{ii}$ were taken directly from MHD simulations plotted
as a function of the phase angle $\phi$.
Each row of panels corresponds to a value of $Q$ and each column to a different
simulation (left: Model A with $10$ times solar rotation; middle: Model B with
20 times solar rotation; right: Model C with 30 times solar rotation).}
\label{fig:input_model}
\end{figure*}

Figure~\ref{fig:input_sin} provides the values of $Q_{xx}$, $Q_{yy}$, and $Q_{zz}$ as a function of the phase angle $\phi$, taking the input as a sinusoidal quadrupole moment. {In these cases,
  the quadrupole moment was constructed by taking the mean value of the MHD model
with rotation rate of $10\Omega_\odot$ (Model A; see Fig.~\ref{fig:input_model}) and
introducing small amplitudes used to create the sine waves.}

Figure~\ref{fig:input_model} shows the input values for the second set of
results corresponding to Fig.~\ref{fig:omc_pencil}, considering model A (10 times solar rotation), model B (20 times solar rotation) and model C (30 times the solar rotation). {A flip in the sign
of the three components of $Q$ can be seen as the rotation increases.
A configuration where $Q_{xx},\,Q_{yy}>0$ ($<0$) and $Q_{zz}<0$ ($>0$) implies an
oblate (prolate) density distribution. $Q_{zz}$ is expected to be
positive as rotation increases. However, we are not including the
centrifugal
force in the MHD simulations. The reason for the change in the sign is due to
increasingly strong magnetic fields that concentrate near the poles of models B and C.}

\end{appendix}

\end{document}